# VIS: the visible imager for *Euclid*

Mark Cropper*[,a], R. Cole [a], A. James [a], Y. Mellier[b], J. Martignac[c], A.-M. di Giorgio [d], S. Paltani [e], L. Genolet [e], J.-J. Fourmond[f], C. Cara[c], J. Amiaux[c], P. Guttridge [a], D. Walton [a], P. Thomas [a], K. Rees [a], P. Pool[g], J. Endicott[g], A. Holland [h], J. Gow [h], N. Murray [h], L. Duvet [i], J.-L. Augueres [c], R. Laureijs [i], P. Gondoin [i], T. Kitching[j], R. Massey[j,k], H. Hoekstra [l] and the *Euclid* collaboration

[a]Mullard Space Science Laboratory, University College London, Holmbury St Mary, Dorking Surrey RH5 6NT, United Kingdom; [b]Institut d'Astrophysique de Paris, 98 bis Boulevard Arago, 75014 Paris, France; [c]Service d'Astrophysique, Commissariat à l'Énergie Atomique, Orme des Merisiers, Bat 709, 91191 Gif sur Yvette, France; [d]Istituto di Astrofisica e Planetologia Spaziali, INAF, Via del Fosso del Cavaliere, 100, 00133 Roma, Italy; [e]ISDC Data Centre for Astrophysics, Chemin d'Ecogia 16, CH-1290 Versoix, Switzerland; [f]Institut d'Astrophysique Spatiale, Campus Universitaire d'Orsay, Batiment 121, Orsay cedex 91405, France; [g]e2v technologies plc, 106 Waterhouse Lane, Chelmsford, Essex CM1 2QU, United Kingdom; [h]Centre for Electronic Imaging, Planetary and Space Sciences Research Institute, The Open University, Walton Hall, Milton Keynes, MK7 6AA, United Kingdom; [i]Research and Scientific Support Department, European Space Research and Technology Centre, Keplerlaan 1, PO Box 299, 2200 AG Noordwijk, The Netherlands; [j]Institute for Astronomy, The University of Edinburgh, Royal Observatory, Blackford Hill, Edinburgh EH9 3HJ, UK; [k]Department of Physics, Durham University, South Road, Durham DH1 3LE, UK; [l]Leiden Observatory, Huygens Laboratory, J.H. Oort Building, Niels Bohrweg 2, NL-2333 CA, Leiden, The Netherlands.

## ABSTRACT

*Euclid*-VIS is a large format visible imager for the ESA *Euclid* space mission in their Cosmic Vision program, scheduled for launch in 2019. Together with the near infrared imaging within the NISP instrument it forms the basis of the weak lensing measurements of *Euclid*. VIS will image in a single *r+i+z* band from 550-900 nm over a field of view of ~0.5 deg$^2$. By combining 4 exposures with a total of 2240 sec, VIS will reach to V=24.5 (10$\sigma$) for sources with extent ~0.3 arcsec. The image sampling is 0.1 arcsec. VIS will provide deep imaging with a tightly controlled and stable point spread function (PSF) over a wide survey area of 15000 deg$^2$ to measure the cosmic shear from nearly 1.5 billion galaxies to high levels of accuracy, from which the cosmological parameters will be measured. In addition, VIS will also provide a legacy imaging dataset with an unprecedented combination of spatial resolution, depth and area covering most of the extra-Galactic sky. Here we will present the results of the study carried out by the *Euclid* Consortium during the *Euclid* Definition phase.

## 1. INTRODUCTION

In the current "concordance model" of the Universe, approximately three quarters consists of dark energy, and approximately one fifth of dark matter. The nature of these constituents is largely unknown. *Euclid*, the second mission in ESA's Cosmic Vision programme, is designed to make the most exquisitely accurate measurements to infer the nature of dark energy, to explore what it is, and to quantify precisely its role in the evolution of the Universe. *Euclid* will additionally measure and elucidate the nature of dark matter. If, instead, the dark energy is a manifestation of a required modification to general relativity on cosmic scales then, *Euclid* will also test the validity of many of these modified gravity theories. Until there is high accuracy data to put these new theoretical frameworks to the test, real progress in constraining the nature of the Cosmos will be limited. *Euclid* will be one of the most powerful tools in this quest, one in which the systematics will be controlled to an unprecedented accuracy through the combination of technical capability and different cosmological approaches.

* msc@mssl.ucl.ac.uk; www.ucl.ac.uk/mssl

Besides these studies in physics and cosmology, *Euclid* will provide a colossal legacy dataset over the whole sky, with optical imaging at 0.2 arcsecond spatial resolution to very faint limits (R~25 at 10$\sigma$), infrared imaging in three bands to similar limits and only slightly worse spatial resolution, and spectra and redshifts of 50 million galaxies to H~19. A dataset of this size will be used by scientists worldwide in a wide range of contexts, and it will have huge scientific and public impact.

This paper discusses the Visible Imager (VIS) on *Euclid*, as envisaged at the end of the Definition Phase (early-2012), updating the position of earlier phases[1,2]. VIS is complemented by the Near Infrared Spectrometer Photometer (NISP) described in a companion papers[3]. The overall scientific aims of *Euclid*, an overview of the mission and the translation to instrumental requirements are also available in these proceedings[4,5] and in the *Euclid* Red book[6].

## 2. PERFORMANCE REQUIREMENTS

### 2.1 Science aims

The main task of VIS is to enable Weak Lensing measurements[7,8,9]. The dark matter (and ordinary matter) aggregates under the influence of gravity as the universe expands. These overdensities distort light from background objects, so that they appear to have an additional, measurable, ellipticity. In general there is only mild distortion: this is weak gravitational lensing. The mass distribution can be reconstructed from the statistical averages of the shapes of background galaxies distorted by this effect, so this is how *Euclid* maps dark matter. Further, by using galaxies further and further away, the characteristics can be determined of the rate at which the agglomeration has occurred. This is directly affected by the expansion history of the Universe, which appears at more recent times to be increasingly driven by dark energy, so the characteristics of the dark energy can consequently also be constrained.

While there is a chain of inferences, in particular the successive removal of the foreground distortions to measure the more distant (and hence from earlier times) distortions, Weak Lensing is considered to be probably the most powerful technique to determine the characteristics of dark matter and dark energy[8,9]. To accomplish this task requires very large surveys in order to ensure a sufficient number of sources and to overcome the natural variations within the Universe, and, also, extremely accurate measurements of galaxy shapes. Systematic effects must be deeply understood and a prerequisite for this is calibrations of the highest quality.

### 2.2 VIS characteristics

VIS therefore requires a large field of view sampled sufficiently finely to measure typical galaxy shapes. To cover most of the extra-Galactic sky in a reasonable mission duration (5–6 years) the field of view must be ~0.5 deg$^2$. To sample galaxies with typical sizes ~0.3 arcsec, pixel sizes of 0.1 arcsec and smaller are required. To minimise the mass of the focal plane and the payload as a whole, the image scale must not be too large, so pixels must be as small, consistent with adequate full well capacity so that there is sufficient dynamic range (otherwise even faint images will saturate) and within a proven technological capability. These requirements are met with a focal plane of 36 CCDs, each of 4kx4k pixels, each 12μm square. We have considered other technologies, in particular infrared arrays such as those in the NISP operated in the visible band, and Active Pixel Sensors, but none of these have the requisite performance characteristics or track record in space for such a large focal plane.

Combined with the NISP infrared photometric measurements (which are however more coarsely sampled) and data from the ground, the Weak Lensing measurements do not require VIS to provide multicolour information within the optical band[10]. Measurements of distant galaxies at blue wavelengths suffer more from the inhomogeneity of galaxies in the ultraviolet (shifted by the expansion of the Universe into the visible band). VIS therefore implements a single broad red band.

Beyond these broad top-level requirements, VIS requires a shutter and calibration unit for flat-fielding the detector, as well as electronics units to process the data from the large focal plane and to control the instrument.

The more general requirements for VIS are given in Table 1.

Table 1: VIS and weak lensing channel characteristics

| Spectral Band | 550 – 900 nm |
|---|---|
| System Point Spread Function size | <0.18 arcsec full width half maximum at 800 nm |
| System PSF ellipticity | <15% using a quadrupole definition |
| Field of View | >0.5 deg$^2$ |
| CCD pixel sampling | 0.1 arcsec |
| Detector cosmetics including cosmic rays | <3% of bad pixels per exposure |
| Linearity post calibration | <0.01% |
| Distortion post calibration | <0.005% |
| Sensitivity | >24.5 at 10σ in 3 exposures for galaxy size 0.3 arcsec |
| Straylight | <20% of the Zodical light background at Ecliptic Poles |
| Survey area | 15000 deg$^2$ over a nominal mission with 85% efficiency |
| Mission duration | 6 years including commissioning |
| Shear systematic bias allocation | additive $\sigma_{sys}$ < 2 x 10$^{-4}$ ; multiplicative < 2 x 10$^{-3}$ |

## 3. EXTERNAL INTERFACES

### 3.1 Optical interface

*Euclid* has a 1.2m 3-mirror telescope, with the third mirror providing corrections to the intermediate Cassegrain focus. The VIS and NISP beams are split by a dichroic, working in reflection for VIS to produce a bandpass of 550 to 900nm with ~20nm band edges. To avoid low-level ghost images from internal reflections, especially as the rays incident on the focal plane are not telecentric, there are no other filters or optics in VIS, except for folding mirrors. The entrance pupil is 1.2m, the central obscuration is 11% and the final focal ratio is f/20.4. This provides an image scale of 0.101 arcsec per 12μm pixel, and a field of view of 0.72°x0.79°, or just under 0.57 square degrees. The optical performance is diffraction limited over the entire field.

### 3.2 Thermal and mechanical

VIS has five assemblies (section 4). Three of the five VIS assemblies, the focal plane, the shutter and the calibration unit, are supported by the payload module structure in the space below the telescope with separate interfaces to the structure. The other two modules are located in the warm service module. Besides the normal requirements for withstanding launch stresses, the positioning of the focal plane with respect to the telescope focus is the most critical mechanical interface.

A cold environment ~150K is provided in the *Euclid* payload module to facilitate the operation of the VIS CCDs, which show optimum charge transfer efficiency at this temperature (see section 6.1). On the other hand, for reasons of device operating conditions, the CCD detection chain electronics must be maintained at a temperature of ≥250K. To provide the two different temperature regimes, two radiators to space are provided by the spacecraft, one for the focal plane and the other for the detection chain electronics.

The total mass of VIS is 102 kg including 20% margin.

### 3.3 Electrical interfaces

The spacecraft Command and Data Management Unit (CDMU) controls VIS through a MIL-STD-1553 bus. Data are transferred to the on-board bulk memory MMU through high speed spacewire links. Primary power (28V, unregulated) is provided by the spacecraft Power Control System (PCDU).

VIS uses 226W of primary power (241W peak) including 20% margin. It produces 400 Gbit of data per day (compressed).

## 4. DESIGN

The five VIS assemblies are shown in Figure 1, with an expanded view of the Focal Plane Array in Figure 2. The electrical architecture is shown in Figure 3.

### 4.1 Focal Plane Array

Photons within the broad red bandpass (550–900 nm) are detected through the array of 36 CCDs, each of which is a CCD273-84[11,12] (Figure 4) specifically designed for VIS and manufactured by e2v Technologies. The CCDs are located in a 6x6 matrix on the front of the Focal Plane Array (FPA). Close butting of the CCDs (Figure 4) which provides a >90% filling fraction of active Si. Optical filters providing a narrower band response previously located over two corner CCDs to calibrate against galaxy colour gradient bias have recently been removed from the baseline.

The FPA has dimensions ~0.45m on each side. The 36 CCDs are held in an Al structure. The detector array is maintained relative to the optical focal plane with tight tolerances to ensure image quality. There is a separate block of electronics which digitise the signals from the CCDs, and is supported separately, with only the CCD electrical flexible interconnects between them (Figure 2). The electronics block holds twelve Readout Electronics (ROE – one for three CCDs), and their closely coupled power supply units (one per ROE). Within the support structure for the ROEs are thermal shields to isolate the cold CCDs from the warm ROEs. One of the two radiators to space maintains the ROE thermal environment, while the other coupled to the detector block maintains the CCDs at a temperature of 150–155K.

The detection chain consists of the CCDs, ROEs and their associated power supply units. The architecture of the chain has been set up to ensure sufficient redundancy, while minimising the use of system resources. Manufacturability, testability and ease of access have also been important considerations. The tradeoff identified the optimal configuration of one ROE supporting three CCDs, each with four video chains (Figure 5), each with its own power supply unit. In order to minimise system noise levels, the power supply for each ROE is accommodated close to it on the outside of the Electronics Block. A single clocking and bias generation circuit is implemented for each CCD. Loss of one ROE results in loss of <10% of the focal plane, and the consequent thermal perturbations are manageable. Within each ROE it is envisaged that individual CCDs may be lost without loss of the entire unit, providing further redundancy.

The CCDs are read out through four readout nodes at a rate of 70 kpix s$^{-1}$. Analogue signals are converted to 16 bits resolution. In order to maintain thermal stability all circuitry remains powered during exposures. Care is taken in the design of mixed signal circuitry handling very low level analogue inputs, to prevent cross-talk and other noise pick-up: multi-layer printed circuit boards are used with separate ground planes for analogue and digital functions. System grounding and decoupling is carefully planned to prevent circulating currents in ground lines from introducing noise sources. All of the CCDs are read out synchronously, minimising radiated and conducted noise during the signal sampling; synchronisation is effected through an LVDS interface (Figure 3). Data are transmitted through a Spacewire port to the payload data handling unit (Figure 3), with Spacewire communication and command decoding, together with other digital functions such as clock sequence generation carried out in a single space-qualified field-programmable gate array (FPGA) per ROE.

Breadboard *Euclid* ROEs have already been designed and fabricated. These have representative board layouts and components, although in standard packaging suitable for use in vacuum equipment, and are currently in use for characterising the CCD273-84 and were also used in the prior testing of the related CCD204-22 and CCD203-84 devices.

The packaging (Figure 5) of the ROE and PSU electronics is designed to optimize conductive thermal paths while minimising the parasitic heat conduction to the CCDs.

### 4.2 Shutter

A shutter is located in front of the FPA to block the light to the detector array while the detectors are not making exposures. This is a momentum-compensated (linear and angular) mechanism in order to minimise any disturbances to the spacecraft and the NISP instrument during actuation. It opens and closes within 10 seconds. The mechanism is electrically cold-redundant and carries a launch lock. The shutter does not seal against any payload module structure, and therefore scattered light paths require careful analysis. Surface treatments for the shutter minimise scattered light when it is opening and closing.

**Focal Plane Array**

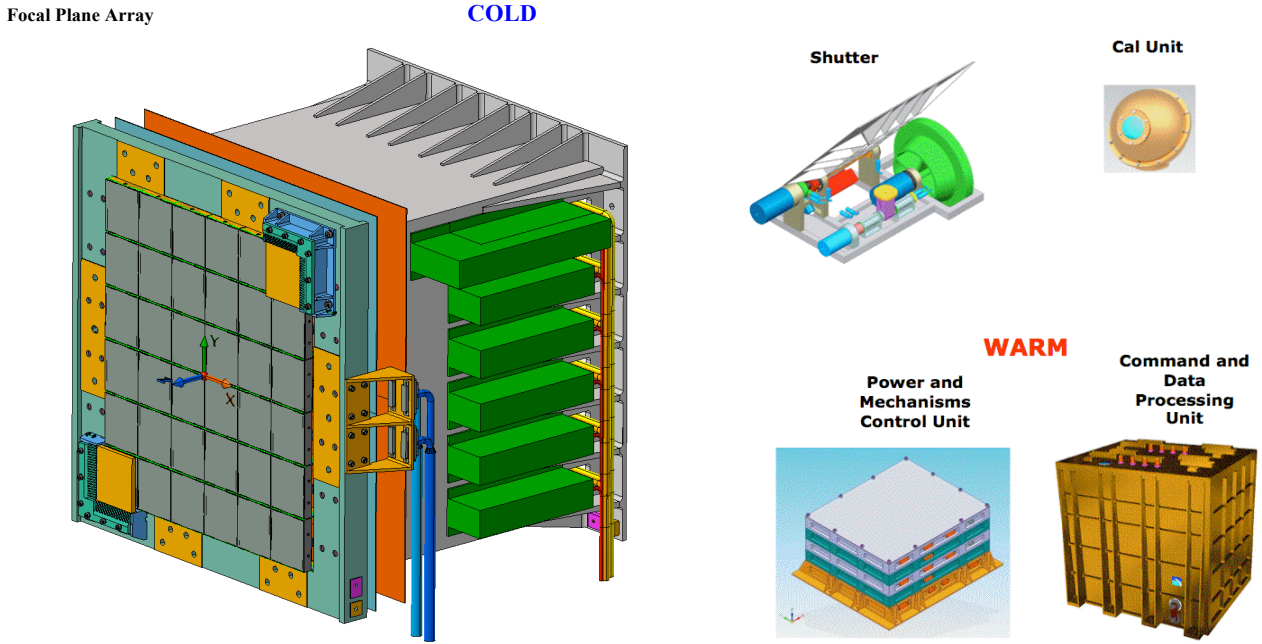

Figure 1: The five units comprising VIS. The two units in the bottom right are in the Service Module and the other three in the Payload Module. The filters over the bottom-left and top-right CCDs are no longer included.

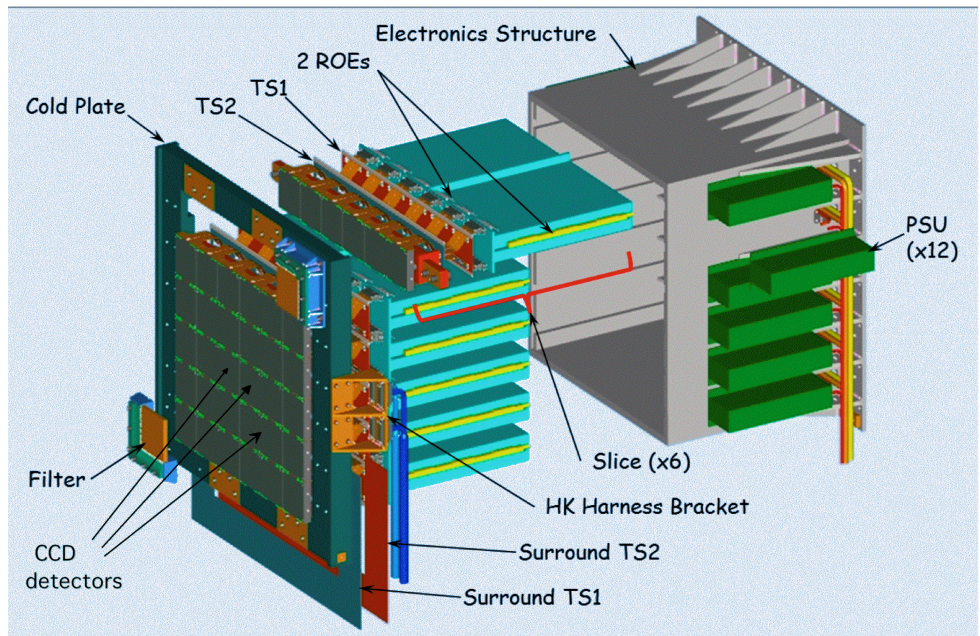

Figure 2: An expanded view of the focal plane array. The detector matrix is on the left, with the top row offset to the rear, to show its relationship of each triplet of detectors to their readout electronics (ROE) which digitise the signals. The cold frame holding the CCDs is mechanically decoupled from the framework holding the ROEs. Each ROE has its own power supply unit (PSU). The filters over the bottom-left and top-right CCDs are no longer included. Items labelled TS are the thermal shrouds to isolate the cold detector plate from the warm ROEs.

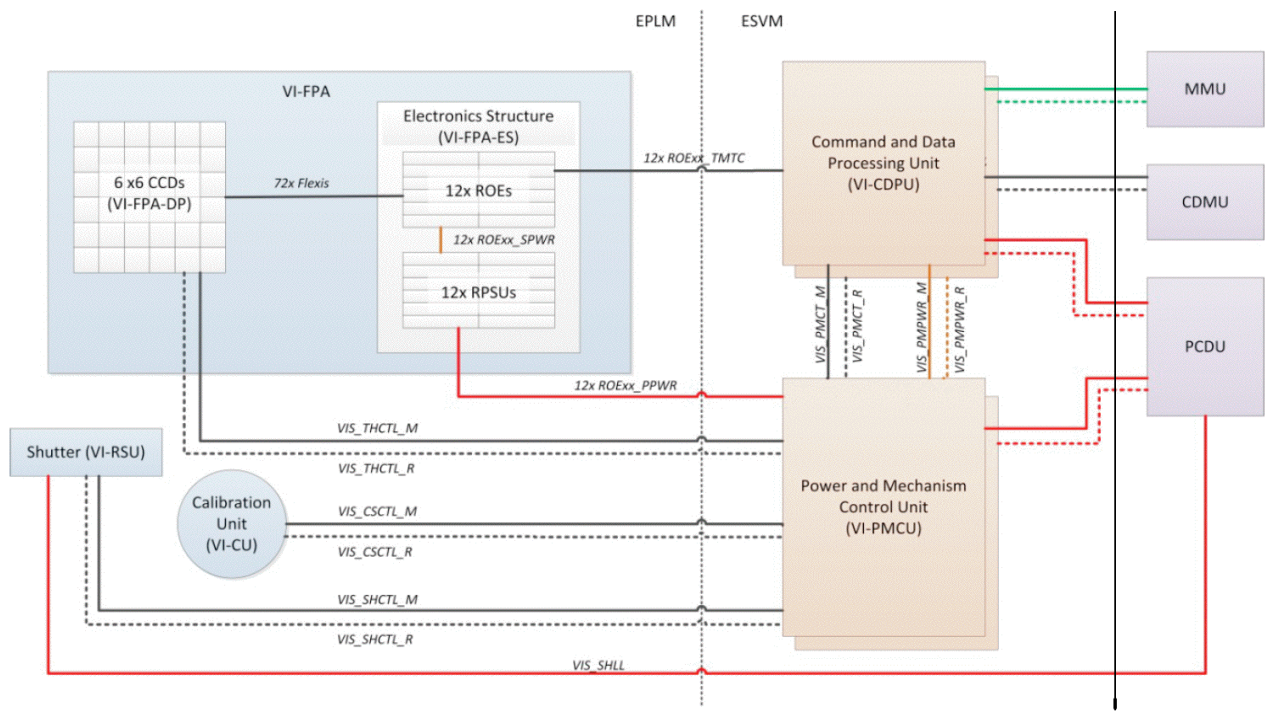

Figure 3: VIS electrical architecture. The three units to the right (MMU, CDMU and PCDU) are spacecraft units.

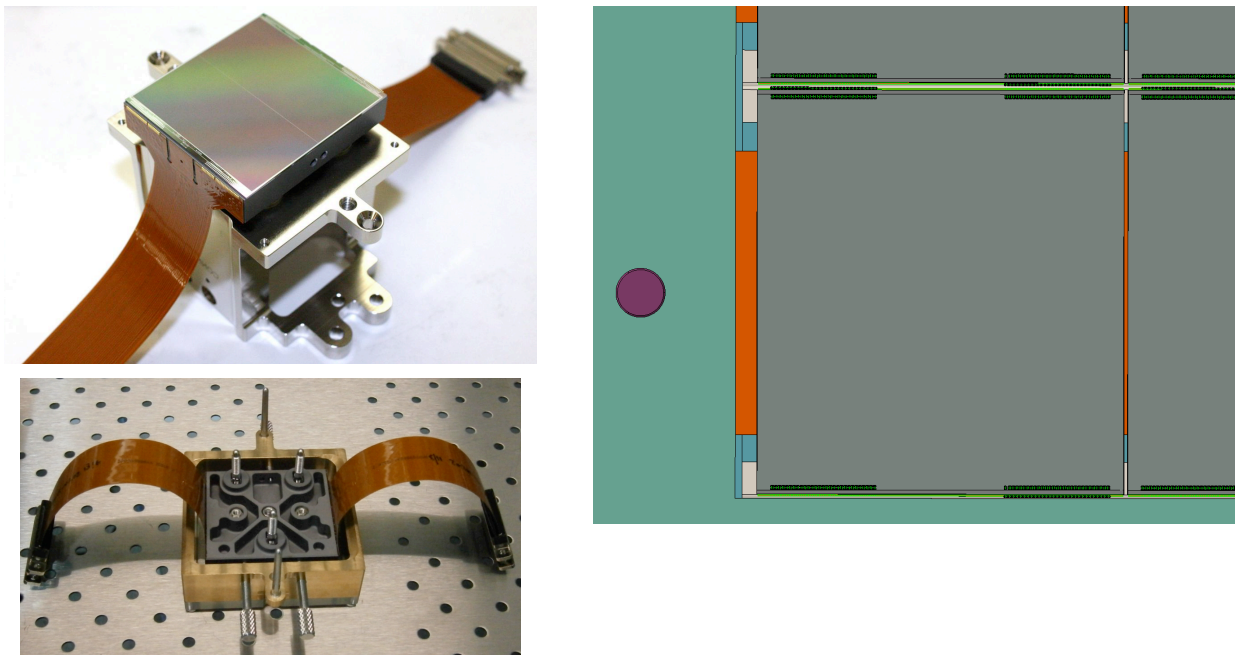

Figure 4: (left) The CCD273-84 to be used in VIS. A view of the active surface (top) shows the arrangement of the flexible connectors to allow close buttability. This is a front-illuminated variant for initial testing, so the charge injection structures are visible down the centre of the device: the devices used in flight will be back-illuminated. The rear surface (bottom) shows the SiC packaging. (Right) The layout of the CCD detector matrix on the FPA.

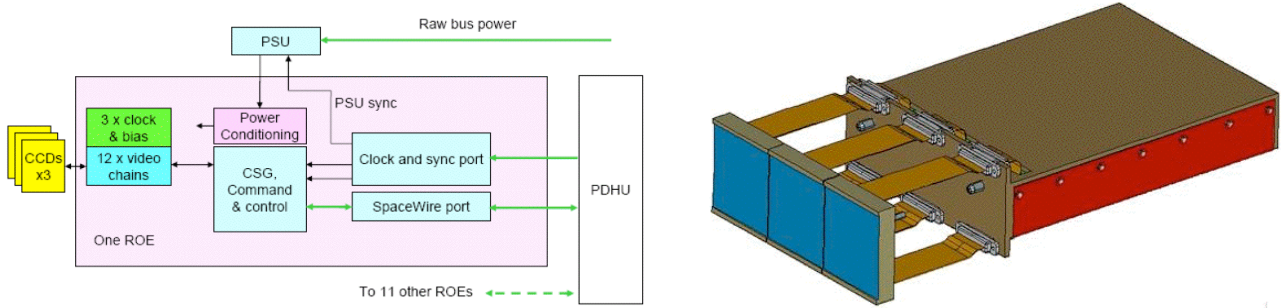

Figure 5: (left) The block diagram for an ROE, interfacing to 3 CCDs. (right) The mechanical packaging of an ROE with 3 CCDs.

### 4.3 Calibration Unit

A calibration unit is provided to flood the FPA with light at three difference wavelengths within the VIS passband. This allows the pixel-pixel variation in the CCDs to be established: it is used for the flat-field calibration. It is not required to provide high levels of stability or of large-scale spatial uniformity, as the photometric throughput is calibrated from stellar sources during routine observations.

There is no shutter for the calibration unit. Flat field calibrations therefore have the sky signal included, but because they are short exposures, and because many flat fields are combined to produce a master flat, this simple solution is acceptable. The calibration unit is electrically cold-redundant.

### 4.4 Command and Data Processing Unit

The CDPU[13] in the service module controls the instrument, transitioning between the different instrument modes and sequencing the operations within exposures, and monitoring instrument status to generate the housekeeping data. It also takes the science data from the ROEs, losslessly-compressing it and buffering it before transfer to the spacecraft bulk memory (MMU). It consists of a science processing unit which performs the compression and science packet generation, a data processing unit to handle the instrument status and the science data from the ROEs.

The CDPU is dual-redundant (Figure 3).

### 4.5 Power and Mechanism Control Unit

The PMCU provides power to the Calibration Unit and the Shutter, and partially conditioned power to the 12 power supplies attached to the ROEs. It also accepts housekeeping data from the shutter and calibration unit, and heater power to the FPA to ensure an appropriate thermal environment. The PMCU is also dual-redundant.

## 5. OPERATIONS AND CALIBRATION

VIS is designed to be simple to operate as this promotes the stability needed for weak lensing measurements. Almost all of the observations are carried out with a 565 second exposure time, with the entire focal plane active (no windowing). The CCDs are then read out, the data are digitised, buffered and compressed, and the process repeated. In order to cover gaps in the detector matrix, to permit some recovery of the spatial resolution, to minimise radiation damage impact on the data and to allow cosmic rays to be identified, four separate exposures are made for each field, with displacements of 100 arcsec between them (with an additional lateral 50 arcsec for the fourth exposure). This results in 50% of the sky being observed in four exposures, and 47% with three, because of the gaps. This sequence is repeated over the area identified for the surveys, eventually covering 15000 deg$^2$ for the Wide Survey and ~40 deg$^2$ for the Deep Survey[5].

The array takes some 80 seconds to read out in total. The VIS exposures take place during the spectroscopic exposures of NISP: these have the same duration. After this, the time during the NISP photometric exposures cannot be used for science observations, because of pointing disturbances caused by the NISP filter wheel. However, they can be used for calibration observations, such as bias fields and dark exposures. Dark exposures will be of three types: in the first the dark current in the detectors (which is extremely low) can be measured, together with hot pixels (which are expected to be relatively few given the cold operational temperature of the CCD); in the second a few lines of charge will be injected electrically into the CCD, and in the third charge will be injected over the whole region of the CCD. These last two allow calibrations to be made of the radiation damage to the detector and are discussed in section 6.1.

Flat field calibrations are made with the Calibration Unit. Because of possible cross-contamination with NISP, these will be done at the end of the sequence of four exposures just before a spacecraft slew. The Calibration Unit is specified to provide sufficient flux within 10 seconds, so this short exposure does not incur any overheads.

Linearity calibrations require a slightly different operational sequence, with different durations for each of the four exposures. In order not to inordinately constrain the shutter repeatability, the shutter will open while the CCD is being read out continuously, after which the readout will stop for a precise duration set by the synchronisation timing in VIS, before the readout begins again and the shutter closes. This will generate stellar images superimposed on illuminated strips in the image, caused by reading out when the shutter is open. These can be subtracted to produce accurate estimates of the flux in the image, and hence to determine the linearity.

Most of the calibrations will however be available from the science data themselves, and in particular the stars on the frame. This includes the photometric calibrations, astrometric calibrations and the calibration of the system PSF (section 6.2 below). This is highly advantageous in that these calibrations are known to be highly representative of the instrument characteristics within each and every exposure.

## 6. PARTICULAR ISSUES

Observations from space permit the fine spatial resolution which is important for weak gravitational lensing measurements of galaxies with size only a small fraction of an arcsecond. In addition, in the absence of an atmosphere, it permits a high level of stability of image characteristics. It is also possible to cover the majority of the extragalactic sky. These characteristics, and the lower sky background allow large surveys with high spatial resolution to be made down to faint magnitudes. *Euclid* VIS will detect more than 1.5 billion galaxies with a signal-noise ratio >10, and this statistical precision afforded by such a large sample is a prerequisite to constraining the nature of dark energy and dark matter. In this case, the particularly difficult aspect of the weak lensing measurements then concerns the control of the statistical biases, given the high precision of the survey. This sets exceptionally tight requirements on the knowledge of the instrumental characteristics, and particularly on the knowledge of the shape of the PSF. Two major issues affect this knowledge.

### 6.1 Radiation

Radiation damage effects in electronics and detectors at the orbit of *Euclid* near the L2 point are largely the result of Solar ions with ~MeV energies. These cause lattice damage in the Si within the pixels and readout registers, creating traps for electrons when the CCD is read out. This leaves a charge trail behind each image as it is moved down the CCD column, and then another in the orthogonal direction as it is transferred along the readout register (Figure 6). The radiation damage effect is parameterised initially through the Charge Transfer Inefficiency (CTI), which measures the fraction of charge lost after the pixel charge contents have been clocked to the readout register, but in *Euclid* a much more detailed characterisation is required in which the details of the charge trail are known to high accuracy.

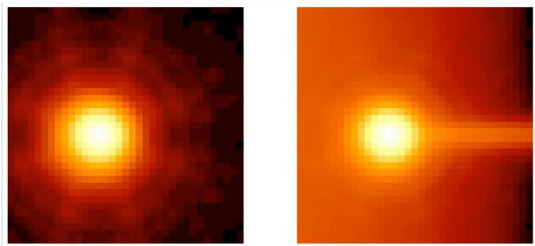

Figure 6: The effect of radiation damage in the CCD. The figure on the left is a simulated source through the *Euclid* system without radiation damage (O. Boulade, CEA Saclay using code from A. Short, ESA) and the right shows the image with radiation damage and cosmic background included when the CCD is read out to the left side of the image. Both are on a log intensity scale to show low-level effects.

These effects are particularly relevant to *Euclid* VIS, because they change the shape of the image, directly affecting the primary quantity to be measured.

**6.2 The VIS point spread function**

The PSF is composed of the optical PSF, the spacecraft pointing jitter and the detector internal charge spread convolved together, and sampled on the CCD pixel grid. The PSF is slightly undersampled by this grid, and some recovery of the spatial frequencies will rely on the combination of the three to four individual exposures.

The main contributor to the PSF is the optics. Given that they are diffraction-limited, the core of the PSF scales nearly inversely with wavelength. There will be some ellipticity in the images resulting from the optical aberrations, and imperfections in manufacture and alignment of the elements. The broadening of the PSF from the spacecraft pointing jitter during an exposure will be different from exposure to exposure, and will depend on the characteristics of the attitude control system. The internal charge spread within the CCDs will depend on their particular operation, and in particular the number of phases held high during the integration. These define the shape of the electrical potential in the column direction, while in the row direction the potential is defined by the column stops. Besides this is the residual effect of the radiation-induced charge trailing, after it has been corrected in the data processing. The shape of the detector-induced contribution to the PSF is therefore slightly non-axisymmetric, which adds ellipticity aligned with the pixel grid to the measured galaxy shapes.

The degree to which the model of the PSF is a faithful representation of the real PSF must be extremely carefully controlled through a detailed budget breakdown which includes all of the instrumental and calibration effects. These must not exceed the additive and multiplicative biases permitted in Table 1. Any model PSF can be constructed by the superposition of different component PSFs, added in the right proportion for each point on the focal plane to achieve a high fidelity model of the actual PSF. In *Euclid* we use principal components analysis (PCA) to generate the components. PCA is a conservative approach: we have demonstrated that with a PCA to generate the components we can achieve the performance required, and, in future, tailored solutions are expected to perform even better. We then fit the components in a Bayesian sense to the stellar profiles measured in the image to determine the proportions of each component. This requires the spectral energy distribution of the stars to be known, and this is available from the combination of VIS and NISP photometry.

## 7. PERFORMANCE

Radiometric modelling predicts that VIS will reach AB magnitude 24.9 ($10\sigma$) in the $r+i+z$ band from three exposures of 565 sec assuming end-of-life conditions and moderately high levels of Zodiacal background. Half of the sources will be covered by four exposures from the dither pattern. Detector noise (mostly readout noise) will be $<4.5e^-$. Stellar sources saturate at $R=17.5$. End-to-end simulations predict that the system PSF will have a full-width-half-maximum 0.17 arcsec. The system ellipticity is ~2%. These indicate that the size and depth of the survey will be sufficient to produce the large number of galaxies required to constrain the cosmological parameters. The largest task in assessing the performance is to quantify the contributions to the biases in the weak lensing measurements, particularly requiring the high level of knowledge of the PSF.

It is clear that radiation damage will be a critical issue for VIS, and a substantial programme is in place to understand and quantify its effect, and to introduce ameliorating measures into the CCD development programme, where these might be possible. CTI correction is limited by the presence of readout noise in the CCD and ROE; a reduction in this has already been achieved. Continuing measures include radiation testing and characterisation, development of software models of varying complexity and of software algorithms for the final data analysis[14]. Many parameters need to be determined for inclusion in the models: in particular the time constants and cross sections of the different trap species, and the beneficial impacts of a non-zero optical background (mostly arising from Zodiacal light) and charge injection have to be quantified. The multi-exposure strategy (dithering) in Section 5 above is an important ingredient in isolating the effect, as each source will be exposed at different locations on the CCD with consequent trails of different extents. The calibration strategy (section 5 again) is designed to provide the information that are required for the models. Full end-to-end performance predictions of continuously increasing fidelity have been developed to quantify its impact. It is important to note however that the effects manifest themselves largely on the angular scale of a CCD quadrant (as there are four readout nodes in the CCD273-84), and not over the full range of scales measured in the 2-point correlation power spectra used to constrain the cosmological parameters.

Once the images have been corrected for the charge trailing and other more standard calibrations applied, the image is available both for legacy science use by the community at large, and for the weak lensing analysis. This analysis requires the effective PSF to be known for each galaxy as described in section 6.2 above. The galaxy ellipticities can then be measured, taking into account this known, locally derived PSF used in the measurement. As currently evaluated, and taking all significant contributions into account, the PSF modelling can be carried out sufficiently accurately to limit the biases to within the allocated limits at the bottom of Table 1 for each image independently. No PSF stability need be assumed from one image to the next, and hence any stability from one exposure to the next provides margin in reaching the required low bias levels. The aggregate of the measured galaxy ellipticities on different spatial scales constrains the weak lensing and hence the cosmological parameters such as the equation of state of dark energy to the required levels[4,6].

## SUMMARY

The *Euclid* VIS instrument, together with the *Euclid* optical system and survey and calibration strategies will produce high spatial resolution images of most of the extragalactic sky, and through measurements of the ellipticities of over $10^9$ galaxies and a tight control of systematic biases, will provide tight constraints on cosmological models. The instrument design is stable and through early development of critical subsystems such as the CCD and ROE, programmatic and technical risks have been minimised. The projected instrument performance has been analysed comprehensively relative to a detailed budget allocation to all of the salient effects: this in itself has required detailed understanding. Current evaluations indicate that all of the science requirements can be met with margin.